\documentclass[twocolumn,prl,superscriptaddress,showpacs,byrevtex]{revtex4}
\usepackage{mathptmx,amssymb,ulem}
\usepackage{amsmath}
\usepackage{graphicx}
\usepackage{color}
\usepackage{epstopdf}

\newcommand*{\pd}{\partial}
\newcommand*{\fb}{\frac}

\newcommand{\be}{\begin{equation}}
\newcommand{\ee}{\end{equation}}

\begin{document}
\title{Dynamo generated by the centrifugal instability}

\author{Florence Marcotte}
\affiliation{Laboratoire de RadioAstronomie, Ecole Normale Sup\'erieure, CNRS, 24 rue Lhomond, 75005 Paris, France}
\affiliation{Institut de Physique du Globe de Paris, CNRS, 1 rue Jussieu, 75005 Paris, France}
\author{Christophe Gissinger}
\affiliation{Laboratoire de Physique Statistique, Ecole Normale Sup\'erieure, CNRS, 24 rue Lhomond, 75005 Paris, France}

\begin{abstract}
We present a new scenario for magnetic field amplification where an electrically conducting fluid is confined in a differentially rotating, spherical shell  with thin aspect-ratio. When the angular momentum sufficiently decreases outwards, an hydrodynamic instability develops in the equatorial region, characterised by pairs of counter-rotating toroidal vortices similar to those observed in cylindrical Couette flow. These spherical Taylor-Couette vortices generate a subcritical dynamo magnetic field dominated by non-axisymmetric components. We show that the critical magnetic Reynolds number seems to reach a constant value at large Reynolds number and that the global rotation can strongly decrease the dynamo onset. Our numerical results are understood within the framework of a simple dynamical system, and we propose a low-dimensional model for subcritical dynamo bifurcations. Implications for both laboratory dynamos and astrophysical magnetic fields are finally discussed.

\end{abstract}

\maketitle

\section*{Introduction} 

During the last decades, satellite observations have significantly improved our knowledge of both the internal structure and the magnetic field of planets in the solar system. Similarly, observations of stellar surface magnetic fields have made important progress in the past few years. For instance, while classical  axial dipole magnetic fields are observed for the Earth, Jupiter or Saturn  (e.g. \cite{LK85,H84,HS84}), data from the Voyager II mission have shown that Neptune's and Uranus' dynamo magnetic fields both exhibit strong non-dipolar components and inclined dipolar field \cite{Ness86,Ness89}. 

These observations have challenged the widely accepted scenario for magnetic  field generation in natural systems, i.e. self-generation of the magnetic field by the turbulent motions of an electrically conducting fluid due to thermo-chemical convection.
For instance, Uranus and Neptune presents similar internal structures, surface magnetic fields and zonal flows, but their measured outward heat flux are very different. Although several studies focused on modifications of convection-driven dynamo models (see for instance \cite{Stanley2004,Aubert2004, Olson10, Gissinger12}), it is therefore tempting to find mechanisms that do not rely on thermal convection but rather depend on other properties, such as planetary internal structure or differential rotation. In this perpective, some authors have studied new types of dynamo forcings likely to occur in planets, such as precession \cite{Tilgner05} or tidal motions \cite{Cebron14}. 

On the other hand, due to the difficulty to generate vigorous convection in experiments, it is currently impossible to reproduce convection-driven dynamos in the laboratory. For this reason, successful dynamo experiments have been obtained so far in setups significantly different from natural systems, in which a liquid metal is pumped inside a constrained pipe geometry \cite{Galaitis01, Stieglitz01} or driven by mechanical impellers \cite{Monchaux06}. More recently, there have been efforts to develop new dynamo experiments based on more realistic flows, such as precession-driven dynamos \cite{Stefani14}. In this perspective, differential rotation may play an important role in astrophysical objects. In planetary cores for instance, shear profiles may be produced by differential rotation between the inner core and outer core-mantle boundaries, which explains why spherical Couette flow, i.e. the flow between two concentric rotating spheres, has been extensively studied in both geophysical and astrophysical framework. 

Recently, the ability of spherical Couette flow to generate a dynamo magnetic field has been studied by \cite{GC10}. They show that in a large gap, the single pair of equatorially symmetric eddies due to secondary meridional circulation in such Couette flows fails to sustain dynamo action in Earth-like geometry (with aspect ratio $\chi=0.35$). Indeed, they find within this geometry that dynamo instability can only build on a first hydrodynamic instability, consisting in the destabilization of the strong equatorial jet, or destabilisation of the quasi-geostrophic shear-layer encompassing the tangent cylinder (the Stewartson $E^{1/4}$ layer, \cite{S66}) when overall rotation is added to the system.

In smaller shell gap, Couette flow may be prone to the centrifugal instability if the angular momentum of the fluid sufficiently decreases outward. There has been considerable interest in characterising such hydrodynamic Taylor-Couette instability of spherical Couette flow -- experimentally \cite{K68,W76}, numerically \cite{B90,MT95,SN01,Y12}, and analytically \cite{HBS00,HBS03}. Although a few number of studies are devoted to the generation of magnetic field by such flow in cylindrical geometry \cite{LCD00,WB02,N12,G14} , the ability of centrifugally-unstable spherical Couette flow to amplify and maintain a magnetic field has never been demonstrated yet. The aim of the present paper is to show that centrifugally-unstable flows are very efficient to produce dynamo magnetic fields in thin spherical shell, and may be relevant for non-convective magnetized systems. Moreover, such spherical Taylor-Couette dynamos display strong analogies with dynamos generated by thermo-chemical convection, although much easier to generate in a laboratory. Their experimental or theoretical study may lead to a better understanding of magnetic field generation in planetary interiors.

The outline of the present article is as follows: we first present our numerical model (section 1) and some hydrodynamic results on the centrifugal instability in a spherical shell (section 2). We then study the Taylor-Vortex dynamo with both non-rotating (section 3) and  rotating (section 4) outer sphere. Finally, we derive in section 5 a simple model based on symmetry arguments which describes the subcritical nature of the dynamo transition.

We conclude with a few remarks on the relevance of this flow for astrophysical and laboratory dynamos.

\section{Model}

We consider an incompressible, electrically conducting fluid confined between two co-rotating spheres.
$r_1$ and $r_2$ are respectively the radius of the inner and outer spheres, and $\Omega_1$, $\Omega_2$ are respectively the angular speed of the inner and the outer spheres. 
The governing equations are the magnetohydrodynamic (MHD) equations, i.e., the Navier-Stokes equation coupled to the induction equation. The problem is made dimensionless following \cite{GC10}: this approach allows us to compare simulations with and without global rotation for a fixed Reynolds number.
\be
\begin{split}
\tilde{Re} \bigg(\fb{\pd u}{\pd t}+ (u \cdot \nabla)u\bigg) + 2 E^{-1} e_z \times u = \\
 - \nabla \Pi + \Delta u + \Big(E^{-1}+\tilde{Re}\chi^{-1}\Big)(\nabla \times B) \times B,
\end{split}
\label{eq1}
\ee
\be
\fb{\pd B}{\pd t}= \nabla \times (u \times B) + \Big(\tilde{Re}Pm\Big)^{-1}\Delta B.
\label{eq2}
\ee

The dimensionless parameters are the Ekman number $E=\fb{\nu}{\Omega_2 r_2^2}$, the Reynolds number $\tilde{Re}=\fb{r_2 r_1 \Delta \Omega}{\nu}$, the magnetic Prandtl number $Pm=\nu/\eta$ and the aspect ratio of the spherical shell $\chi=r_1/r_2$, where $\nu$ and $\eta$ are respectively the kinematic viscosity and the magnetic diffusivity. The inner and outer spheres are electrical insulators, and no-slip boundary conditions are used on these boundaries. In all the results reported in this paper, a thin gap $\chi=0.9$ is used. Note that the Reynolds number used here as a control parameter ($\tilde{Re}$) is based on the outer shell radius. In the following, our results will be presented in terms of a more conventional Reynolds number ($Re$) based on the gap width $\delta=(r_2-r_1)/r_2)$ to allow easier comparison with previous works: $Re=\fb{\delta r_1 \Delta \Omega}{\nu}$.
The equations (\ref{eq1})-(\ref{eq2}) are integrated using the PARODY code \cite{D97}. This code relies on a toroidal/poloidal decomposition of the velocity and magnetic fields. It uses spherical harmonics expansion in the latitudinal and azimuthal directions for each spherical shell, whereas finite differences on a stretched grid are used in the radial direction. Time integration is performed using a Crank-Nicholson scheme for the diffusive terms and an Adams-Bashforth scheme for the nonlinear terms. Typical numerical resolutions involve  $64$ grid points in the radial direction, while retaining $196$ spherical harmonics in the latitudinal direction ($l_{max}=196$), and $96$ in the azimuthal direction ($m_{max}=96$). This resolution has been proved sufficient in the covered parameters range by convergence tests (up to $l_{max}=256$, $m_{max}=128$, and $96$ radial gridpoints).

\section{Centrifugal Instability in spherical shell}

In the idealized situation where the spheres are replaced by two infinitely long rotating cylinders, the ideal laminar Couette solution for the rotation $\Omega$ is given by :
\begin{equation}
\Omega(r)=A_1+\frac{A_2}{r^2},
\end{equation}
in which $A_1=(\Omega_2 r_2^2-\Omega_1 r_1^2)/(r_2^2-r_1^2)$ and $A_2=r_1^2 r_2^2(\Omega_1-\Omega_2)/(r_2^2-r_1^2)$. In this case, the 
Rayleigh criterion predicts linear stability if the angular momentum increases outwards ($\Omega_1 r_1^2 \le \Omega_2 r_2^2$). Otherwise, the centrifugal instability is observed when the Reynolds number is large enough, and takes the form (at onset) of axisymmetric Taylor-Couette vortices contained in the poloidal plane.
A similar behavior occurs in a spherical gap: if the angular momentum decreases outwards along a cylindrical radius, Taylor vortices can be generated near the equator if the gap is small enough. Note that in this case, the bifurcation is strongly imperfect, since it appears from a background flow already involving a poloidal recirculation. Nevertheless, for a sufficiently thin spherical shell, one can expect Rayleigh's criterion to remain valid at least close to the equatorial plane, which translates with our dimensionless numbers to the condition:
\be
\label{Rayleigh2}
E \ge \left(\chi^{-1} -\chi \right)\delta Re^{-1}.
\ee
\begin{figure*}[htb!]
\begin{center}
 \includegraphics[width=0.7\textwidth]{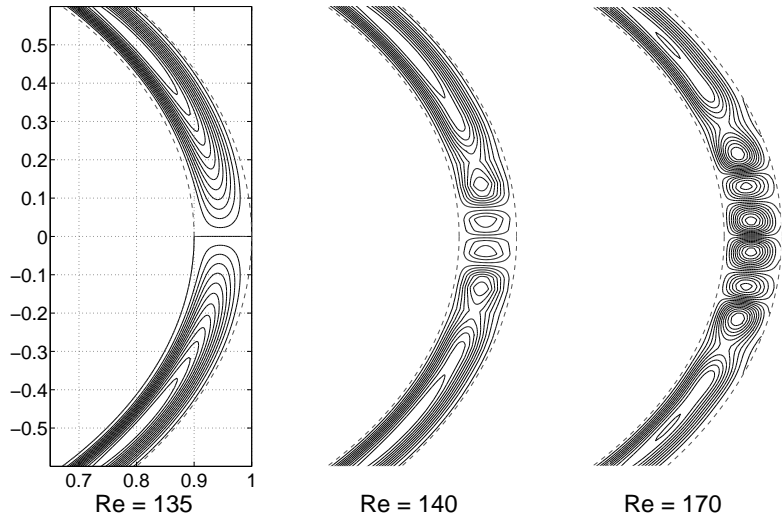}
\caption{Streamlines of the azimuthally averaged velocity field, for different Reynolds number below ($Re=135$, i.e. $\tilde{Re}=1350$) and above ($Re=140 ; 170$, i.e. $\tilde{Re}= 1400 ; 1700$) the Taylor Vortices threshold, without overall rotation.}
 \label{fig:stream_Einfini}
 \end{center}
\end{figure*}

Let us first study the case with no global rotation, i.e. when $\Omega_2=0$. At low Reynolds number, the mean flow is equatorially symmetric and essentially toroidal, except for a large-scale meridional recirculation between polar and equatorial regions (Fig.\ref{fig:stream_Einfini}-a ). However, characterisation of the equatorial flow in the narrow-gap spherical shell indicates the existence of a Rayleigh-like stability threshold, above which several pairs of counter-rotating, toroidal and axisymmetric Taylor vortices (TV) build up on the primary base flow. The new base state above this threshold corresponds to the saturated TV concentrated in the vicinity of the equator, as shown in Fig.\ref{fig:stream_Einfini} for several Reynolds numbers. This bifurcation has been widely studied in different aspect-ratios by \cite{B90,MT95,SN01,Y12}.

Several purely hydrodynamic simulations have been performed for Reynolds numbers up to $\tilde{Re}=4500$  ($Re=450$) in order to characterise the successive bifurcations of this spherical Couette flow. The domain-integrated kinetic energy of the poloidal velocity field is plotted against the Reynolds number in Fig.\ref{fig:bifurcTV}-a showing an imperfect, supercritical bifurcation at approximately $Re=140$. This critical value is in good agreement with both experimental results by \cite{K68} (predicting $Re_c=147$ for $\chi=0.9$) and more recent numerical estimations \cite{Y12} (linear extrapolation from their results at $\chi=0.877$ and $\chi=0.84$ yields $Re_c \sim 144$). From $Re_c=140$ up to approximately $Re=170$, the TV are purely axisymmetric, while subject at larger $Re$  to secondary hydrodynamic instabilities: then they break into wavy vortices, characterised by a dominant azimuthal mode $m=4$ at $Re=170$ and $m=5$ at $Re=180$. As the flow becomes more chaotic at larger $Re$, all modes become unstable, thus yielding a complicated azimuthal structure of the Taylor Vortices which become less and less coherent.
 
\begin{figure}[htb!]
\begin{center}
\includegraphics[width=0.5\textwidth]{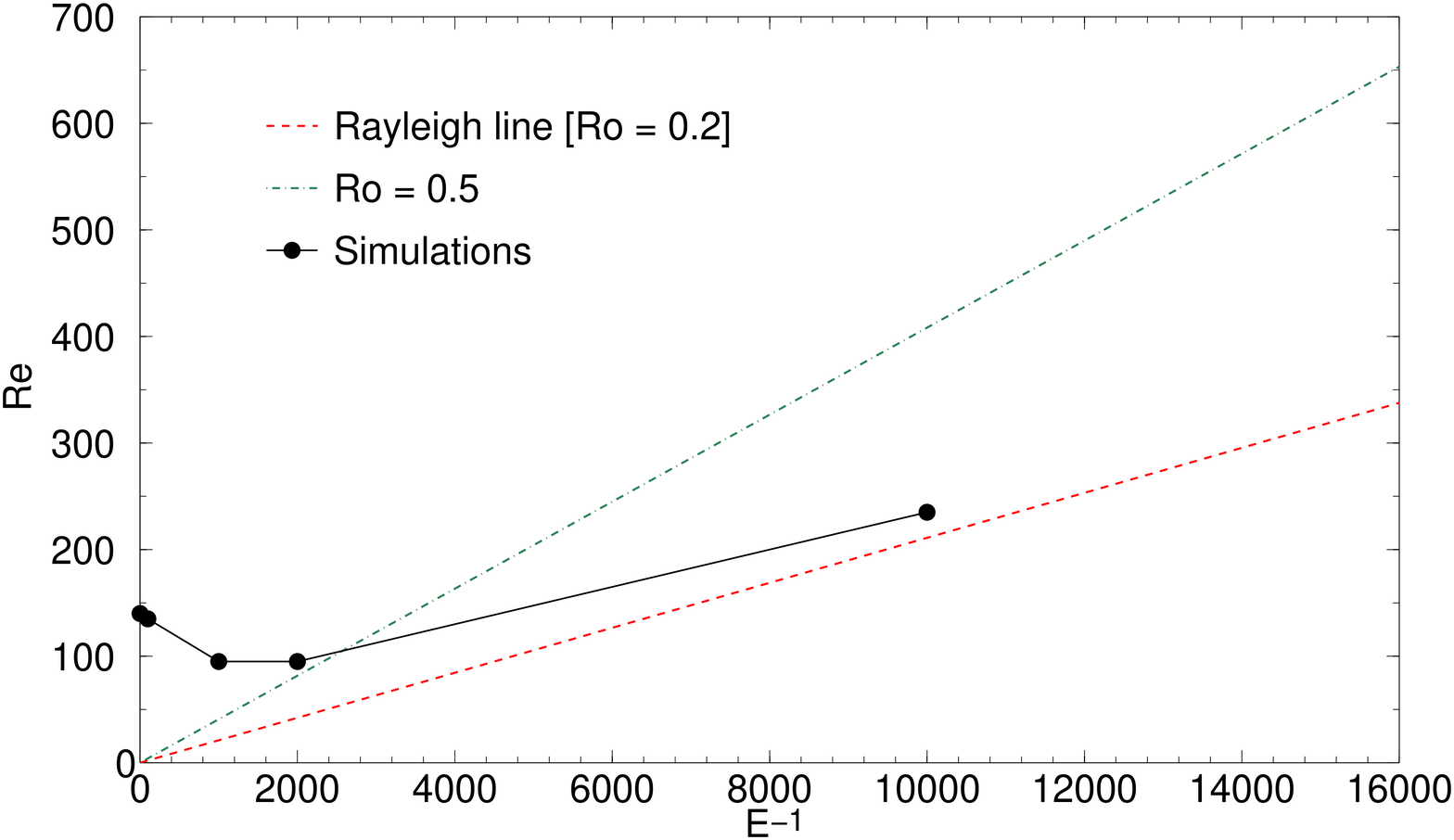}
\caption{Marginal stability curve for the destabilization of laminar Couette flow in a spherical gap, for the purely hydrodynamic problem. Red line indicate Rayleigh's criterion for infinite cylinders (\ref{Rayleigh2}).}
\label{fig:RecE}
\end{center}
\end{figure}

We now consider the case where, adding overall rotation to the system, the Coriolis force is taken into account in the momentum equations (expressed in the frame of reference of the outer sphere). For different Ekman numbers ranging from  $E=10^{-2}$ up to $E=10^{-4}$, we systematically performed purely hydrodynamic simulations in order to determine the onset of TV and the structure of the new base state up to weakly turbulent Reynolds numbers. These results are illustrated in Fig.\ref{fig:RecE}, which shows the marginal stability line for pure spherical Couette flow in the $E^{-1}-Re$ plane, showing a non-monotic evolution of the onset with the inverse Ekman number.

In addition, the domain of existence of axisymmetric vortices significantly varies with the overall rotation rate. For $E=10^{-3}$ for example, axisymmetric TV are to be observed up to at least $Re=300$, where they break into wavy TV ($m=6$), while axisymmetry is lost from $Re=170$ already ($m=4$) in the $E=\infty$ case. An extensive hydrodynamic study is far beyond the scope of the present paper ; nevertheless the Fig.1 in \cite{ALS86}, although in cylindrical geometry, provides a glimpse into the complexity of the TV phase diagram in spherical Couette flow.

Below the TV threshold, the kinetic energy contained in the two meridional recirculation cells increases quasi-linearly with $Re$ as can be seen in Fig.\ref{fig:bifurcTV}-a. We therefore approximate this contribution with a linear fit and subtract it from the total kinetic energy associated with poloidal motion close to the TV threshold. 
In Fig.\ref{fig:bifurcTV}-b we present this last quantity as a function of the supercriticality coefficient $Ta/Ta_c$, where $Ta$ is the Taylor number ($Ta_c$ the TV critical Taylor number), as defined in \cite{C61}: $Ta=\fb{4\Omega_1^2}{\nu^2}{R_1}^4\fb{(1-\Omega_2/\Omega_1)(1-\chi^{-2}\Omega_2/\Omega_1)}{(1-\chi^2)^2}$. This has the advantage of providing a single dimensionless parameter to describe the TV bifurcation, for all overall rotation rates.

\begin{figure}[htb!]
\begin{center}
\includegraphics[width=0.45\textwidth]{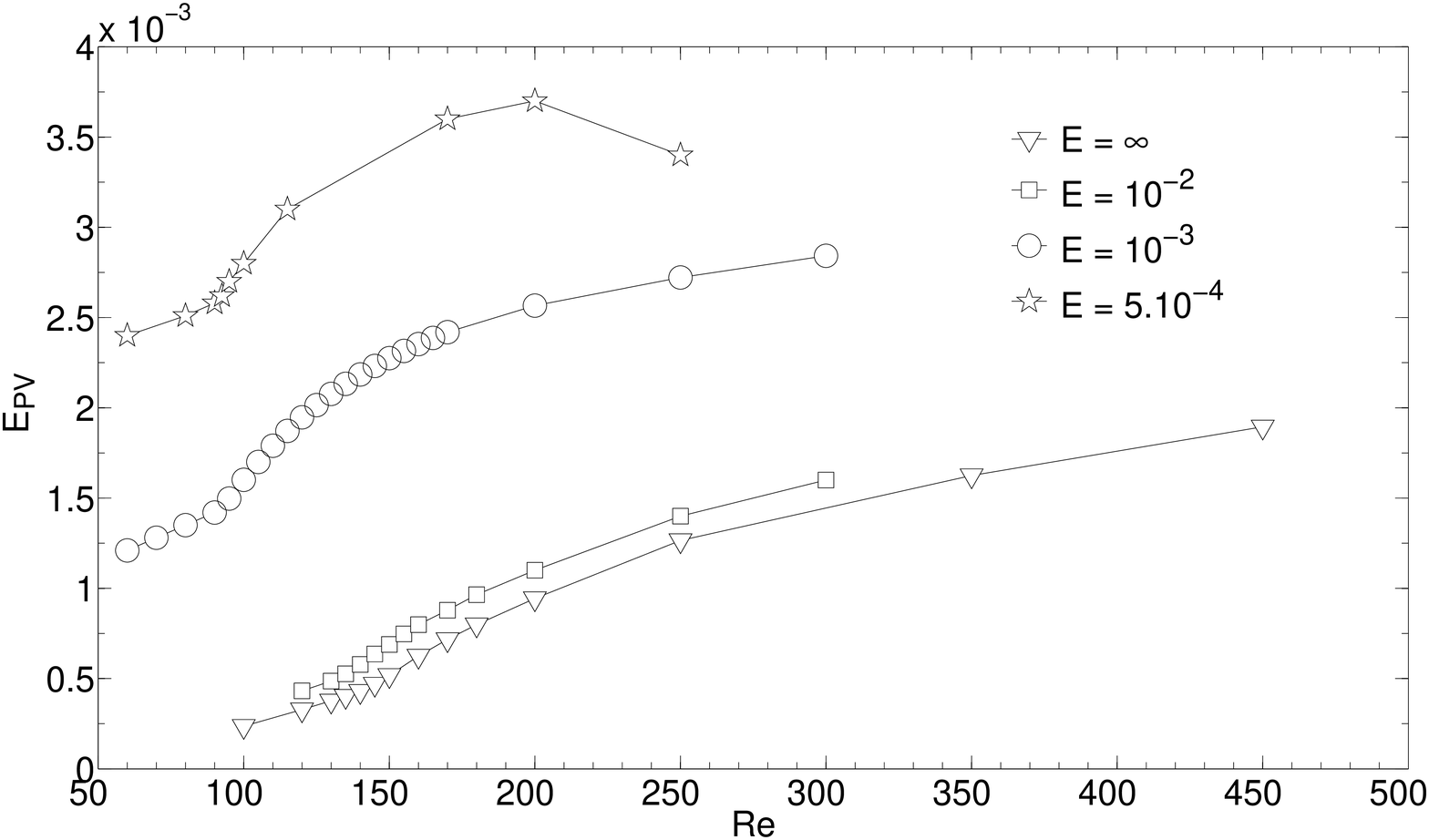}
\includegraphics[width=0.45\textwidth]{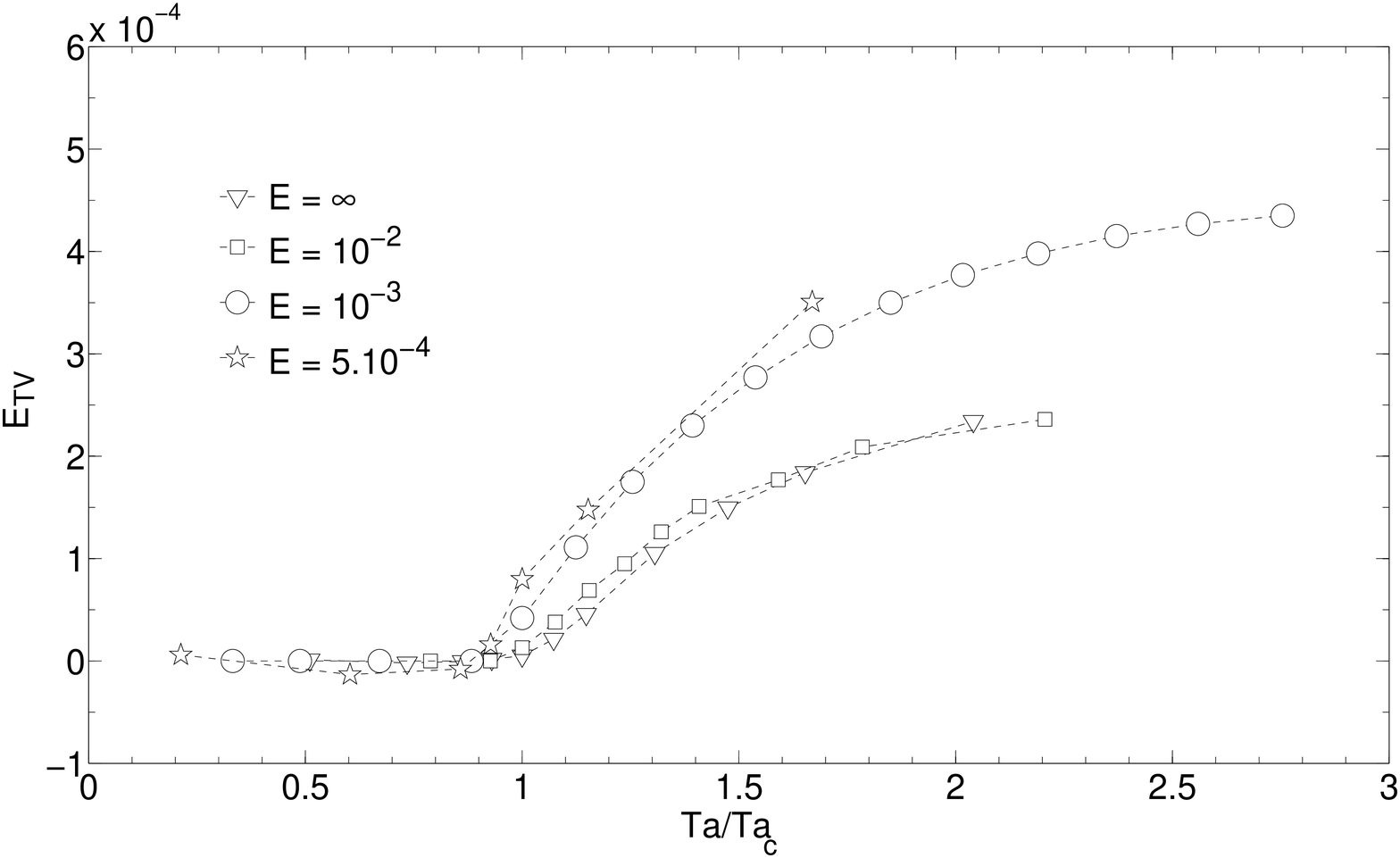}
\caption{\textit{\textbf{Top:}} Evolution of the domain-integrated kinetic energy associated with poloidal motion ($E_{PV}$) as a function of the Reynolds number, for four different Ekman numbers ($E=\infty$, $E=10^{-2}$, $E=10^{-3}$ and $E=5.10^{-4}$). \textit{\textbf{Bottom:}} Kinetic energy contained in the Taylor Vortices ($E_{TV}$, see text) as a function of the distance $Ta/Ta_c$ to the TV onset, for the same Ekman numbers (retaining only weakly supercritical points).}
\label{fig:bifurcTV}
\end{center}
\end{figure}

\section{Taylor vortex dynamo at infinite Ekman number}

\begin{figure}[htb!]
\begin{center}
\includegraphics[width=0.48\textwidth]{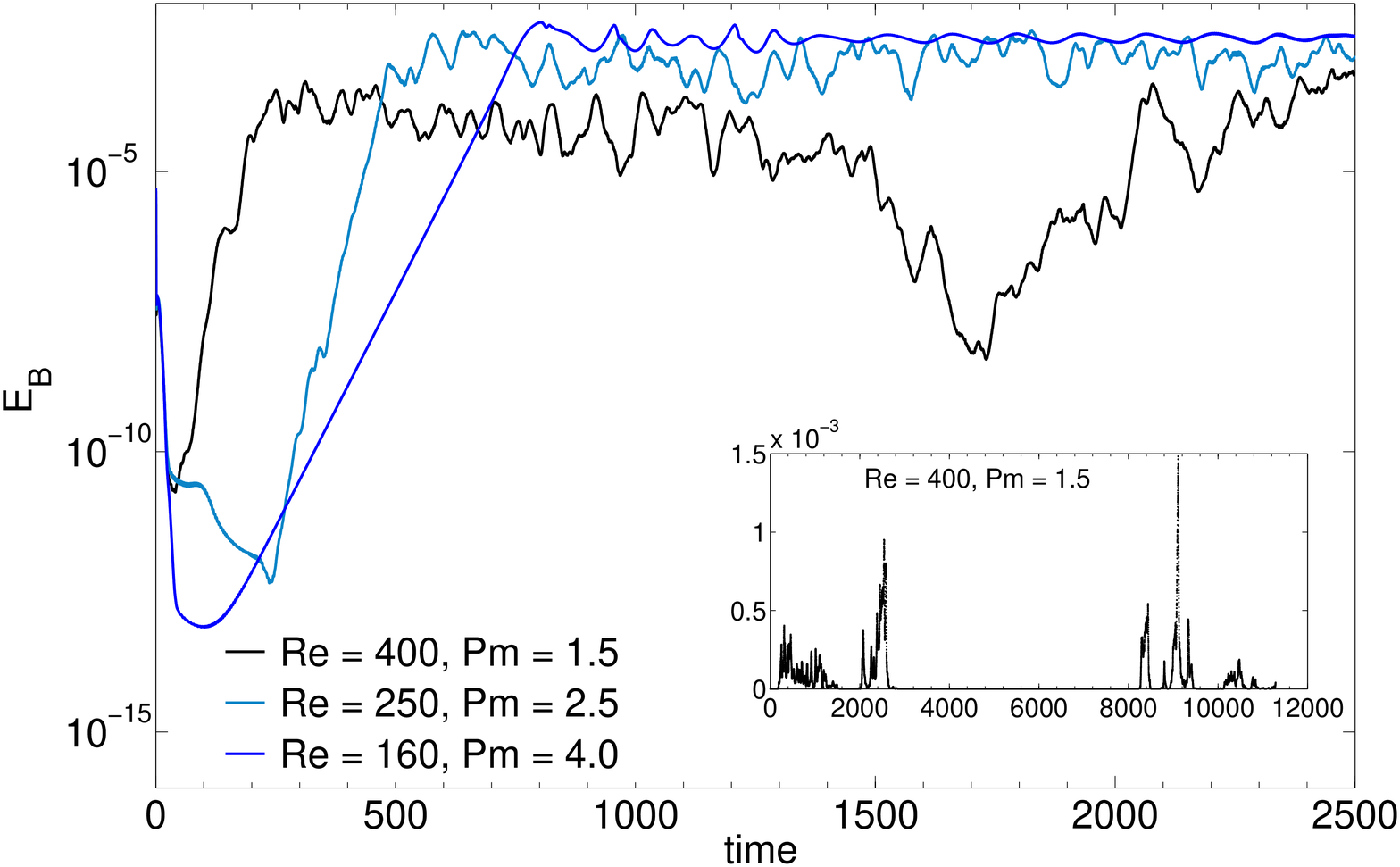}
\caption{Time series of the domain-integrated magnetic energy for $\{Re=160, Rm=640\}$ , $\{Re=250, Rm=625\}$ and $\{Re=400, Rm=600\}$.  For larger $Re$, the magnetic field exhibits intermittent bursts near the dynamo threshold (inset). }
\label{fig:series_temp}
\end{center}
\end{figure}

In this section, we first assume the outer sphere to be at rest ($\Omega_2=0$) and investigate the possibility for a dynamo instability to arise in the absence of Coriolis force. Solving now for the full set of MHD equations (\ref{eq1}-\ref{eq2}), the flow is found to sustain a dynamo in the equatorial region as soon as the bifurcation for axisymmetric TV is reached. For sufficiently high values of the magnetic Prandtl number $Pm$, a seed magnetic field experiences an exponential growth followed by a saturated phase as the feedback of the Lorentz force modifies the base flow. Fig.\ref{fig:series_temp} shows the time evolution of the magnetic energy for different values of the Reynolds number. First, one can see that the kinematic growth rate of the magnetic energy increases with the fluid Reynolds number, although the magnetic Reynolds number $Rm=PmRe$ slightly decreases. Note that this behavior is quite different from what is observed in the cylindrical geometry, where the turbulent fluctuations decrease the growth rate of the magnetic field, such that dynamo action is inhibited at large $Re$ (see Fig.4 in \cite{G14}).

The dynamics in the saturated phase also strongly depends on the level of turbulent fluctuations. At small $Re$, both the magnetic energy and the kinetic energy undergo smooth periodic oscillations around a well defined value. As $Re$ is increased, the magnetic field exhibits more and more chaotic behavior under the effect of turbulent fluctuations in the TV. Finally, for larger $Re$, the magnetic field exhibits intermittent growth with short, noisy saturation phases when close to the dynamo onset (see inset of Fig.\ref{fig:series_temp}).

\begin{figure}[htb!]
\begin{center}
\includegraphics[width=0.48\textwidth]{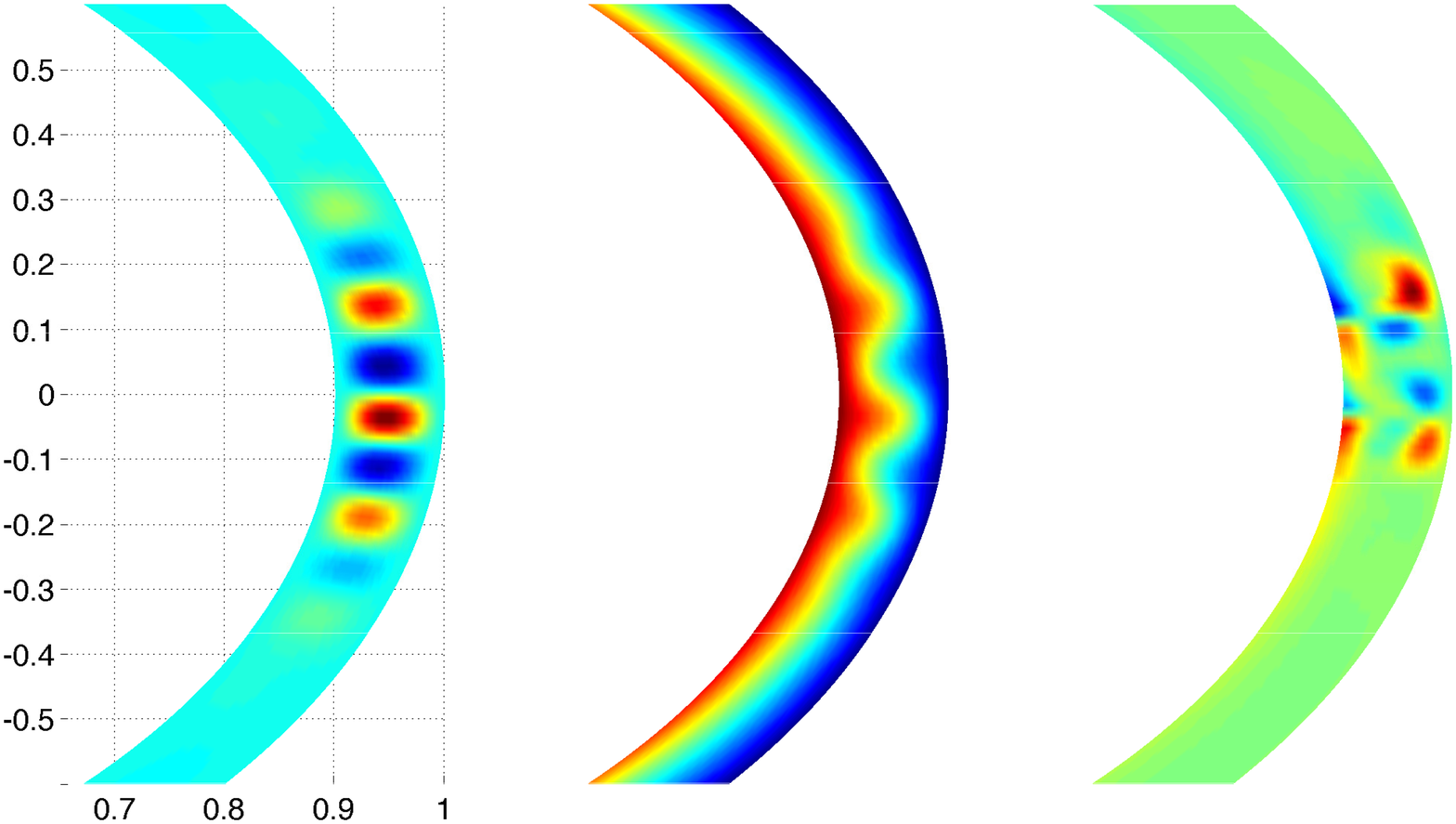}\\
\includegraphics[width=0.25\textwidth,angle=90]{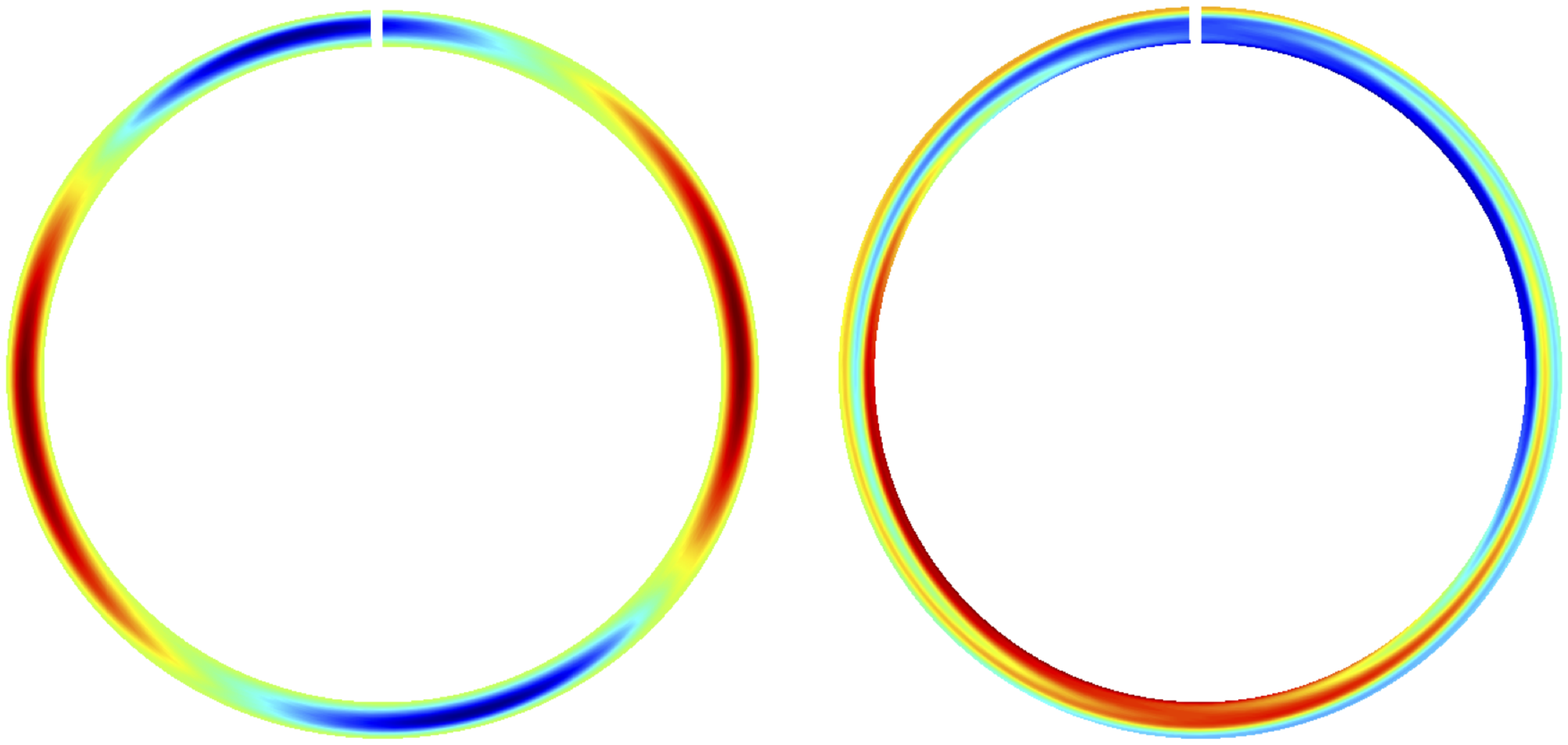}
\caption{The $m=1$ dynamo at $Re=150$, $Pm=7$. \textit{\textbf{Top, from Left to Right:}} Meridonal maps of radial velocity $V_r$, azimuthal velocity $V_{\phi}$, and latitudinal magnetic field $B_{\theta}$, \textit{\textbf{Bottom, from Left to Right:}} Non-axisymmetric part of the radial components of $V$ and $B$, in the equatorial plane.}
\label{fig:mapRe1600PM4iEk0}
\end{center}
\end{figure}

For $Re=150$, Fig.\ref{fig:mapRe1600PM4iEk0}-top shows velocity components $V_r$, $V_\phi$, and magnetic component $B_\theta$  in a meridional cut ($\phi=0$) , whereas Fig.\ref{fig:mapRe1600PM4iEk0}-bottom shows non-axisymmetric radial component of velocity and magnetic fields in the midplane $\theta=0$. This figure illustrates the non-dipolar character of the dynamo field in which   the dominant mode  near the onset is an azimuthal mode $m=1$, which in turn sustains a $m=2$ structure in the velocity field through the action of the Lorentz force. The longitudinal structure of the magnetic field is essentially determined by those of the forcing TV: while always subharmonic in the cylindrical geometry \cite{G14}, the spherical Taylor-Couette dynamo displays both harmonic and subharmonic features in the vertical direction.

\begin{figure}[htb!]
\begin{center}
\includegraphics[width=0.48\textwidth]{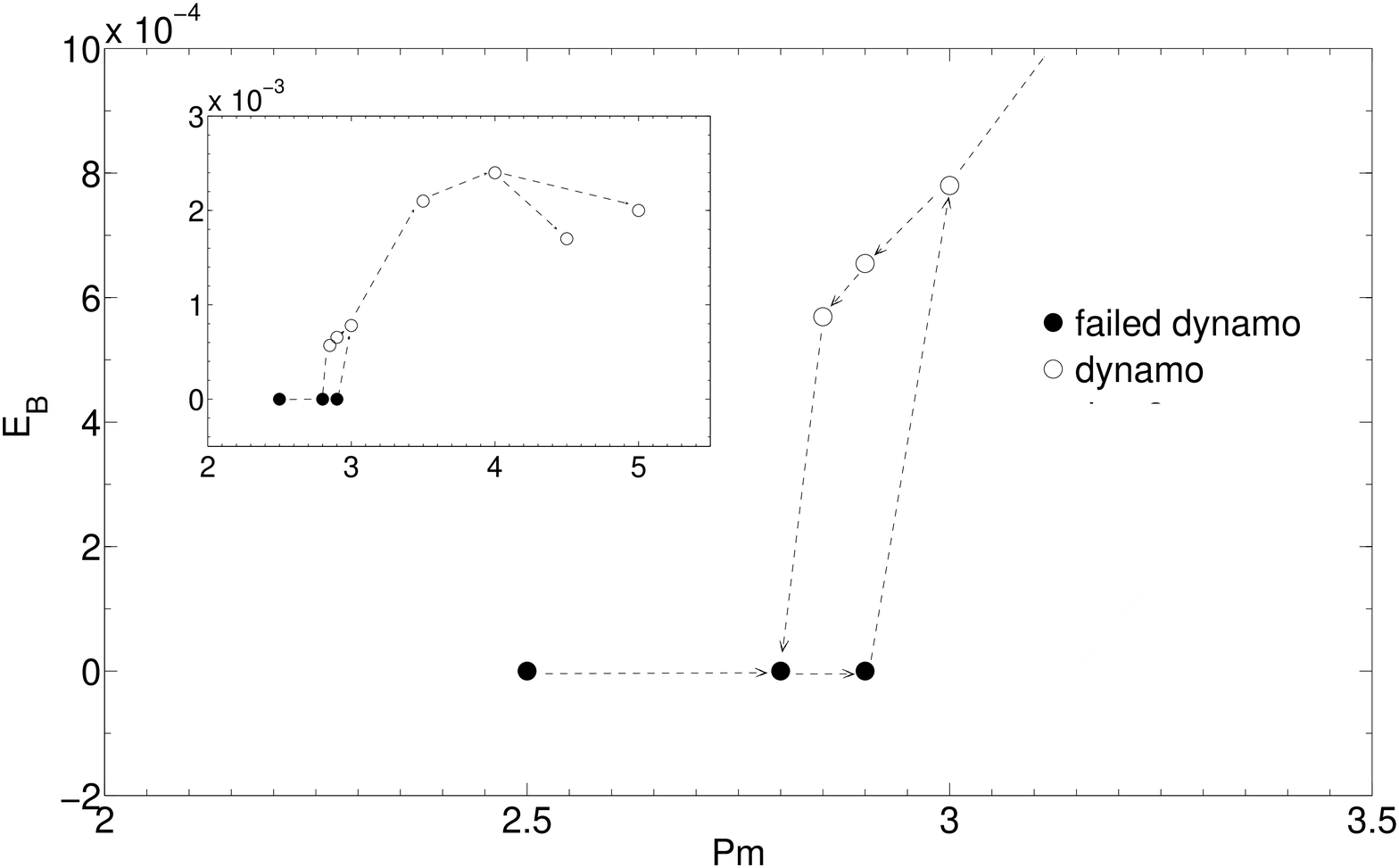}
\includegraphics[width=0.48\textwidth]{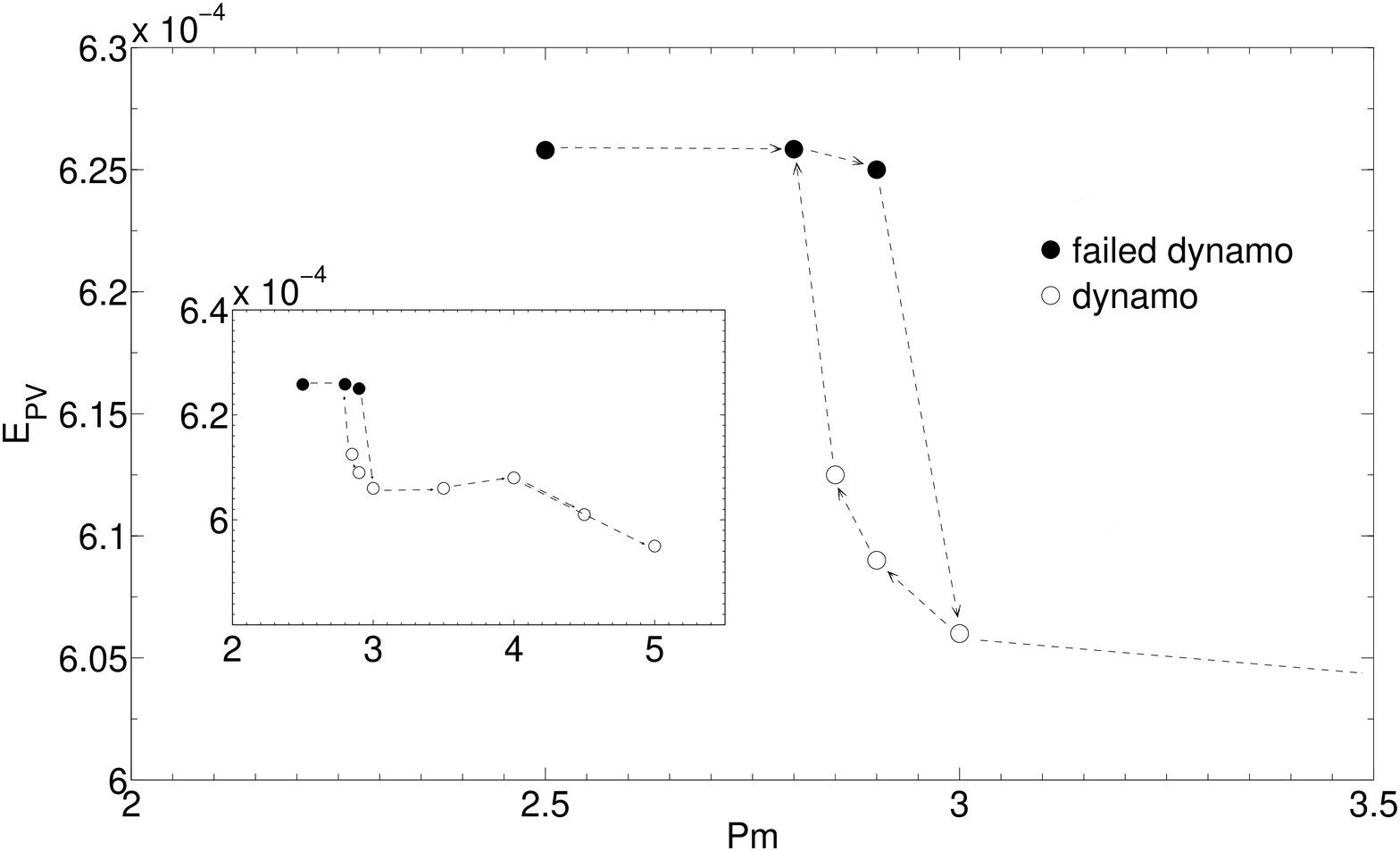}
\caption{Bifurcation diagram of domain-integrated magnetic \textit{\textbf{(top)}} and poloidal kinetic \textit{\textbf{(bottom)}} energy vs $Pm$, for $Re=160$ and $E=\infty$. For these parameters, the dynamo bifurcation is clearly subcritical.}
\label{fig:bifurcRe1600}
\end{center}
\end{figure}

The bifurcation diagram of Fig.\ref{fig:bifurcRe1600}-a displays the evolution  of the time-averaged, integrated magnetic energy after saturation of the dynamo, as a function of the magnetic Prandtl number at a fixed value of the Reynolds number $Re=160$. Fig.\ref{fig:bifurcRe1600}-(b) shows the corresponding time-averaged, integrated kinetic energy contained in poloidal motion after saturation. The bifurcation is clearly subcritical and displays well-defined hysteresis near the dynamo threshold, the amplitude of the TV weakening in concert with the strengthening dynamo as the flow transfers kinetic energy to the magnetic field. It is interesting to note that the hysteresis is smaller close to the TV onset $Re_c$ and tends to a supercritical transition in this case. As $Re$ is increased, the magnetic dominant mode switches to $m=2$ (for $Re\ge160$), triggering a $m=4$ component in the velocity field.

\begin{figure}[htb!]
\begin{center}
\includegraphics[width=0.5\textwidth]{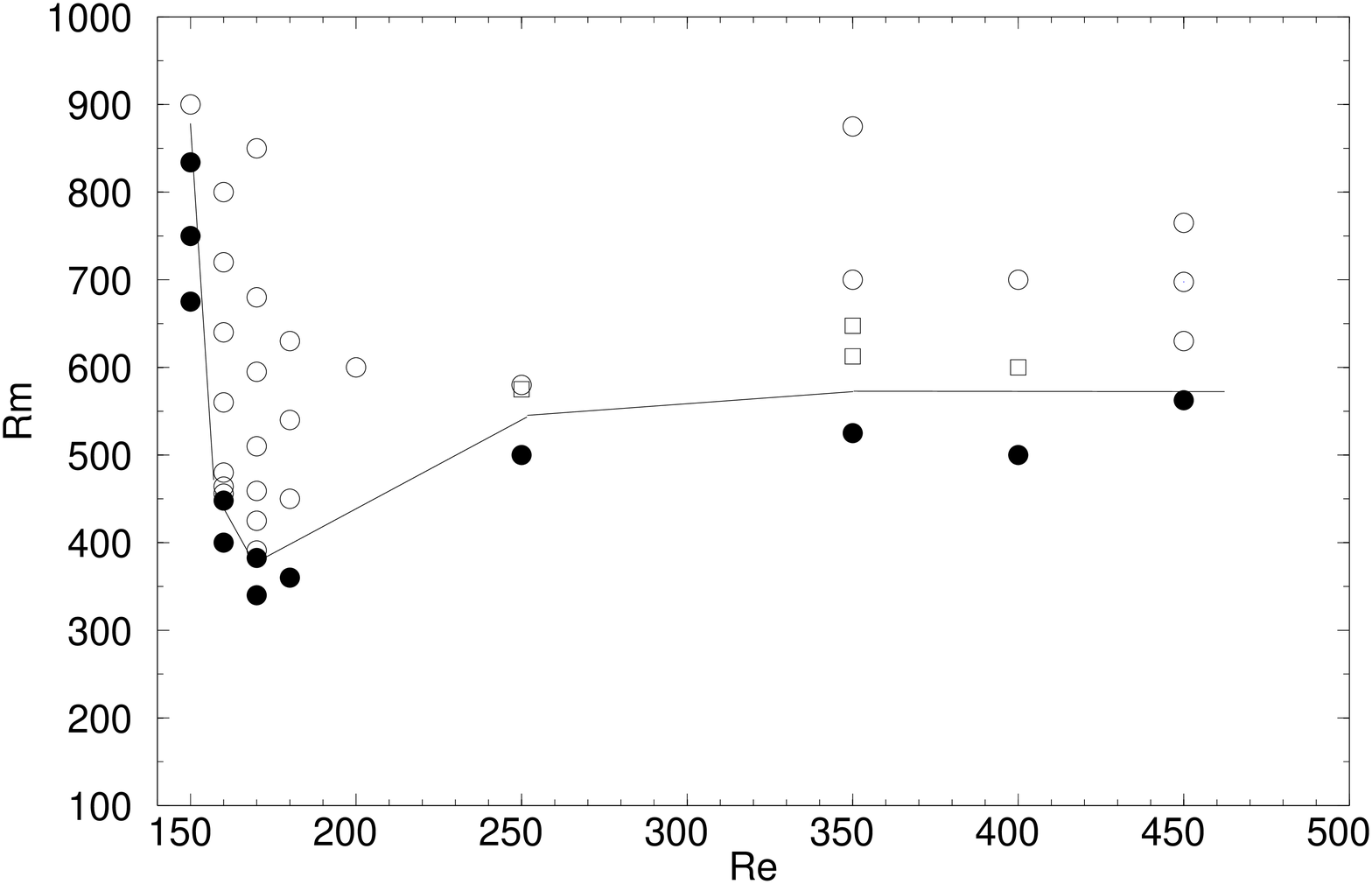}
\caption{Magnetic Reynolds number $Rm$ versus $Re$, without overall rotation of the system. Empty circles denote dynamo runs (whether subcritical or not); filled circles: failed dynamos, and squares: intermittent bursts. Note the divergence at low $Re$ and the constant critical $Rm$ at large $Re$. }
\label{fig:Rm_vs_Re}
\end{center}
\end{figure}

In Fig.\ref{fig:Rm_vs_Re}, we report the marginal stability curve for dynamo action in the $(Rm-Re)$ parameter space.  First, the critical magnetic Reynolds number $Rm_c$ for magnetic field amplification clearly diverges near the hydrodynamic TV onset at $Re_c=140$ , meaning that the threshold for dynamo action is essentially determined by the onset of TV, without which no magnetic field can be sustained by the flow. Note however that this $Rm_c$ diverges only very close to $Re_c$, but reaches its minimum value immediately above this point ($Re_c^{min}\sim 170$). This is similar to what is observed for convective dynamos, for which convection instability is necessary, or to the results obtained by \cite{GC10}, who demonstrated in a large-gap geometry that the single pair of equatorially symmetric eddies due to secondary meridional circulation could not, as such, sustain a dynamo in spherical Couette flow. The most favorable window for dynamo action (in terms of critical $Rm$) seems to lie in the $Re=170-180$ range. Interestingly, this window also corresponds to the regime of wavy, but still laminar Taylor Vortices. As we will see in section 4, this last feature is essential for the comprehension of the dynamo bifurcation.

It is remarkable that the critical magnetic Reynolds number for dynamo action $Rm_c=Pm_cRe$ seems to saturate above $Re=350$. Indeed, the key element for the spherical Taylor-Couette dynamo is the coherence of the TV,  crucial in sustaining the magnetic field. In the cylindrical case (see \cite{G14}), further increase in Reynolds number results in a loss of coherence in the TV as the flow develops chaotic features, which results in a loss of efficiency for the dynamo. A similar behavior is observed in large gap spherical Couette dynamo or weakly rotating convection dynamos, for which a critical magnetic Prandlt number $Pm_c$ is obtained at large $Re$, rather than our critical $Rm$.

Numerical resolution requirements did not allow us to reach values of $Re$ above $450$ (corresponding to our control parameter $\tilde{Re}=4500$). However, recent work by \cite{Y12} on the hydrodynamics of spherical Couette flow in relatively thin gaps up to large Reynolds number (corresponding to $Re\sim 850$ with our definition) shows a re-laminarisation of the TV.

\section{Effect of global rotation on the dynamo}

In this section, we now investigate the effect of the rotation of the outer sphere on the dynamo, which is equivalent to study the effect of global rotation on the Taylor-Vortex dynamo described in the previous section. We have computed  hydrodynamic and MHD simulations for different values of $E=\{10^{-1};10^{-2};10^{-3};5.10^{-4} \}$ in order to obtain both TV onset and dynamo threshold. 

\begin{figure}[htb!]
\begin{center}
\includegraphics[width=0.5\textwidth]{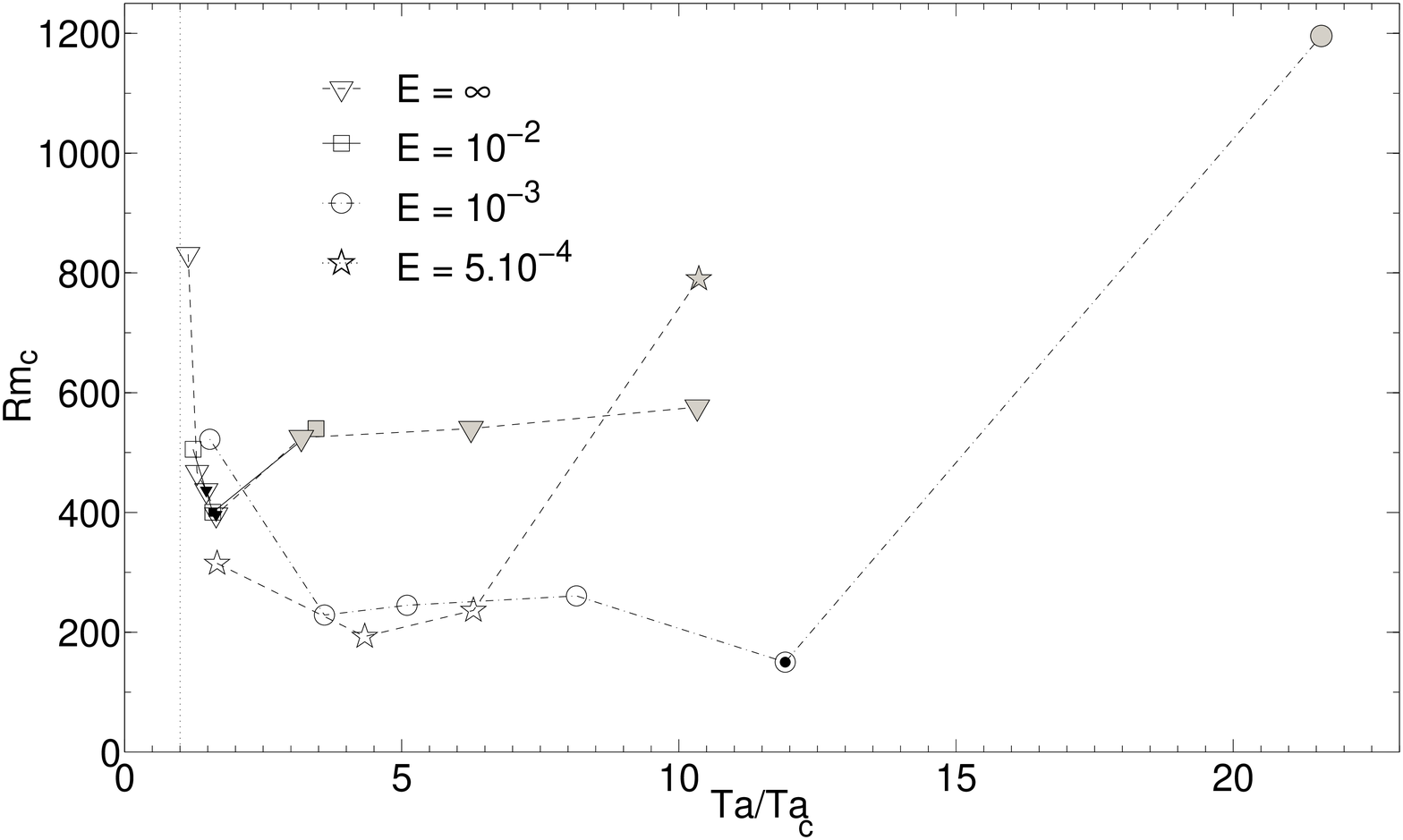}
\caption{Critical $Rm$ as a function of $Ta/Ta_c$ (where $Ta_c$ is the TV critical Taylor number), for different values of the Ekman number $E$. Empty markers denote laminar, axisymmetric TV (in the purely hydrodynamic regime). Markers filled in black denote laminar, wavy TV and markers filled in grey correspond to turbulent flows.}
\label{fig:Rmc}
\end{center}
\end{figure}

We first compare in Fig.\ref{fig:Rmc} the dynamo threshold - in terms of the critical magnetic Reynolds number $Rm$ - for various global rotation rates, as a function of the supercriticality coefficient $Ta/Ta_c$ (for TV onset).  Note that our marginal curve corresponds to the linear instability, i.e. critical $Pm$ is obtained by interpolation of time-averaged kinematic growth rate of the total magnetic energy computed close to the dynamo onset. 

For all the parameters explored so far, the dynamo threshold diverges very sharply close to the onset of TV, such that the loss of TV vortices is always accompanied with the loss of dynamo action, similarly to the non-rotating case (Fig.\ref{fig:Rm_vs_Re}).
Although there are some uncertainties related to the estimation of both the critical $Re$ for TV onset and the critical $Pm$ for dynamo,   the following picture emerges from the simulations: for all the overall rotation rates investigated here, the dynamo threshold decreases as the Taylor number is gradually increased from its critical value $Ta_c$, while still in the laminar, axisymmetric TV regime. It further drops as the hydrodynamic base state switches from axisymmetric to wavy TV, reaching its minimum value in this new regime. Then, once the weakly turbulent regime is reached, the dynamo threshold rapildy rises to a higher value, which seems nearly constant in the infinite Ekman number case. (We could not investigate sufficiently high $\tilde{Re}$ regime to confirm this last observation for smaller Ekman numbers). 

Close to the TV threshold or in the turbulent regime, overall rotation of the system does not particularly help dynamo action. However, at reasonable distance from the TV onset, the linear, critical $Rm$ can be decreased down to considerably lower values (by nearly a factor 3) when the system is rapidly rotating. It is interesting to compare the flow realizing the minimum $Rm_c$ for each overall rotation rate investigated here.

Interestingly, the minimal dynamo onset is systematically obtained in the vicinity of the hydrodynamic instability to wavy vortices: for $E=\infty$ and $E=10^{-2}$, the dynamo is characterised by a $m_B=2$ azimuthal wavenumber and exhibits a minimum onset for $Ta/Ta_c \sim 1.5$, while wavy vortices $m_V=2m_B=4$ are hydrodynamically unstable for $Ta/Ta_c \sim 1.35$. Similarly, for $E=10^{-3}$, we observe a minimum onset at $Ta/Ta_c \sim 12$, with the most unstable mode $m_B=3$, very close to the onset for waviness $m_V=2m_B=6$ at $Ta/Ta_c \sim 11$.


From Fig.\ref{fig:Rmc} it is clear that the apparently favorable influence of overall rotation on dynamo action is not simply determined by the distance to the TV onset. Moreover, the strength of the hydrodynamic forcing by the TV does not explain the discrepancy in the minimal thresholds. 
We computed an effective magnetic Reynolds number $Rm^e_c$, based on the gap width and the TV forcing velocity, which allows us to distinguish between the contribution of the large-scale, non-dynamo Couette flow and the r\^ole of the equatorial TV forcing responsible for the dynamo action at scale $\delta$. We find $Rm^e_c=49.8$ for $\{E=10^{-3},Ta/Ta_c=11.92\}$ against, for example, $Rm^e_c=107.7$ for $\{E=10^{-2},Ta/Ta_c=1.59\}$. This example illustrates well that overall rotation is capable of effectively favouring dynamo action, independently from the hydrodynamic forcing strength.


\section{Low-dimensional model for the subcritical dynamo bifurcation}

We now present a simple model aiming to explain how a subcritical dynamo bifurcation can be generated in such system, and why the minimum threshold is systematically obtained close to the transition to wavy Taylor vortices. For the parameter $\{ Re, Rm\}$ explored in these simulations, the system is always close to both the dynamo transition and the wavy vortex bifurcation, leading to a codimension-two  bifurcation. We denote by $A(t)$ the complex amplitude of the non-axisymmetric magnetic field with wavenumber $m_B$, its phase describing the angle of the magnetic mode in the equatorial plane. This non-axisymmetric mode generates a non-axisymmetric velocity field of complex amplitude V(t) through the Lorentz force, which depends quadratically on the magnetic field. When writing amplitude equations for these two modes, symmetry requirements constrain the form of the equations: $A\rightarrow -A$ (symmetry of the induction equation) and $A\rightarrow Ae^{i\Psi}$, $V\rightarrow Ve^{2i\Psi}$ (rotational invariance about the rotation axis), which yields:
\begin{align}
\label{mode2m}
\dot{V}&= \lambda V + A^2 - |V|^2 V,\\
\label{modem}
\dot{A}&=  \sigma A + V \bar{A}- |A|^2 A,
\end{align}

The quadratic terms $A^2$ and $V\bar{A}$, which respectively represent the Lorentz force and the induction term of MHD equations, describe the coupling between the dynamo mode with wavenumber $m_B$ and the wavy component $m_V=2m_B$  of the flow. $\lambda$ represents the linear growth rate (or equivalently the distance from onset) of the wavy mode in a non-magnetized situation ($A=0$), whereas $\sigma$ is the linear growth rate of the dynamo in absence of wavy vortices  $m_V$. Note that the effect of axisymmetric TVs, main source of the dynamo, is implicitly described by $\sigma$. At leading order, we have therefore $\lambda \sim Re-Re_c^W$ and $\sigma \sim Rm-Rm_c$, where $Rm_c$ and $Re_c^W$ are respectively the onset for dynamo action in the absence of waviness in the flow and the onset for resonant wavy vortices (in the absence of magnetic field). The above model is the normal form of the well known $1:2$ resonance and has been studied in various situations (\cite{AGH88,Gissinger08}). Let us write $A=r e^{i\theta}$, $V=se^{i\phi}$, and $\psi=\phi-2\theta$ and look for steady solution ($\dot{r}=\dot{s}=\dot{\psi}=0$). The model has two trivial solutions : $\{s=0,r=0\}$ and the pure {\it hydrodynamic} mode $\{s=\sqrt{\lambda},r=0\}$. In addition, the mixed modes of the model are given by the equations:
\begin{align}
s&= \left(r^2 - \sigma \right),\\
0&=\left(\lambda \sigma - \sigma^3 \right) + \left(3 \sigma^2 - \lambda - 1\right) r^2 -  3 \sigma r^4 + r^6.
\end{align}
This low-dimensional model provides a simple explanation for the subcritical nature of the dynamo. Let us consider $\lambda<0$, i.e. hydrodynamically damped wavy vortices: for $\sigma>0$, the solution $A,V=0$ is linearly unstable and bifurcates to a mixed mode $A\ne0$, associated to a velocity mode $V\ne0$ due to the action of the Lorentz force. Far away from the onset of wavy vortices $|\lambda | \gg 1$, it is easy to show that the equation for $A$ reduces to a quadratic equation, meaning that the dynamo bifurcation is supercritical.

\begin{figure}[htb!]
\begin{center}
\includegraphics[width=0.5\textwidth]{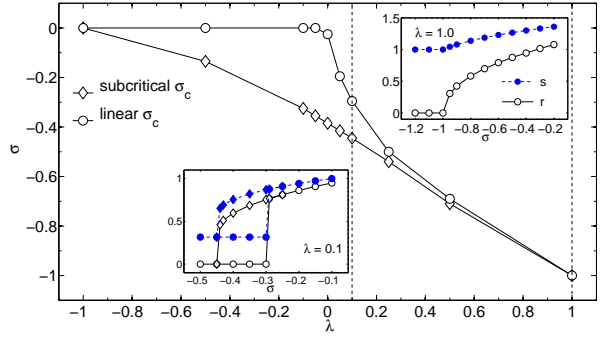}
\caption{Stability diagram in the $\lambda-\sigma$ parameter space, where both subcritical and linear thresholds $\sigma_c$ are represented for each $\lambda$. \textit{\textbf{Insets:}} Subcritical (respectively, supercritical) bifurcation diagrams for $\lambda=0.1$ (respectively, $\lambda=1.0$).}
\label{fig:siglam}
\end{center}
\end{figure}
\begin{figure}[htb!]
\begin{center}
\includegraphics[width=0.5\textwidth]{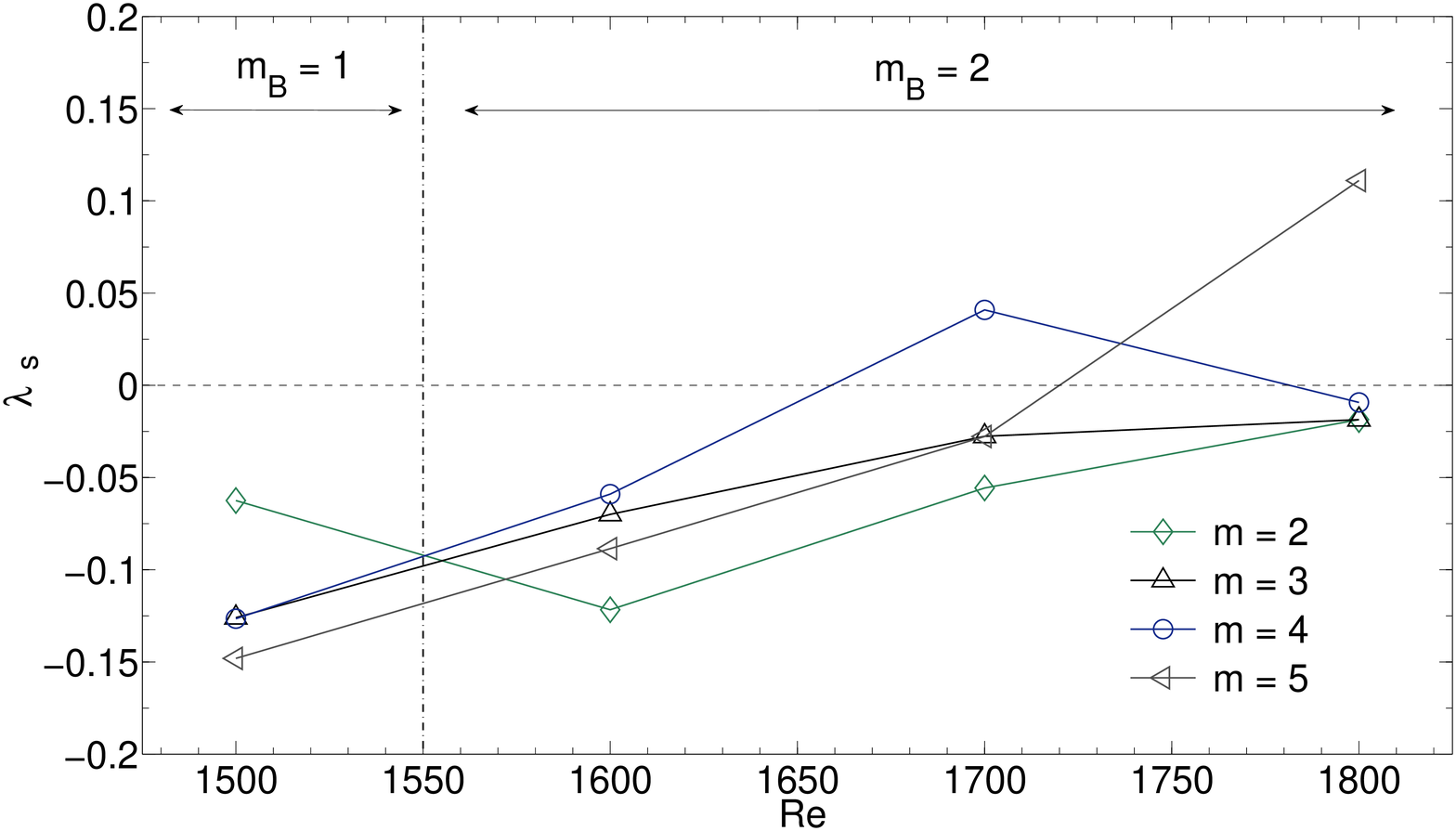}
\caption{Growth rates ($\lambda_S$) in the simulations of the hydrodynamic wavy modes $m=2-5$ at $E=\infty$. For comparison, the magnetic mode $m_B$ selected at the dynamo onset is indicated. As can be seen in Fig.\ref{fig:Rm_vs_Re}, the minimal $Rm_c$ lies in the vicinity of $Re \sim 170$, corresponding to the hydrodynamic bifurcation of wavy vortices.}
\label{fig:lambdam}
\end{center}
\end{figure}

On the contrary, very close to the onset for hydrodynamic wavyness, $|\lambda | \ll 1$ leads to a subcritical transition: the mixed mode can be stabilized and is therefore bistable with $A,V=0$ for $\sigma<0$. This mechanism is illustrated in Fig.\ref{fig:siglam}, where the model (\ref{mode2m})-(\ref{modem}) has been integrated in time for various values of $\sigma$ and $\lambda$, and for various initial conditions (but such that $\psi=0$ is preferred over $\psi=\pi$). As $\lambda$ increases from large, negative values (where the dynamo bifurcation is supercritical) to zero, the subcritical threshold for the dynamo lowers, resulting in a widening bistability region. Note that in the vicinity of $\lambda=0$, subcritical bifurcation is obtained independently of the sign of $\lambda$. Then, for $\lambda> 0$, as the hydrodynamic mode becomes more and more linearly unstable, the linear dynamo threshold drops faster than the subcritical threshold, which results in recovering a supercritical bifurcation at large $\lambda$ (see upper inset in Fig.\ref{fig:siglam}).

This model shares several similarities with the equations derived in \cite{K11}, who recently used a dynamical system to show that a competition between an hydrodynamic instability and a growing magnetic mode can lead to a subcritical dynamo. Note however that the 1:2 resonance underlying the present approach directly constrain the symmetry of the modes that can be involved in the bifurcation, and the quadratic order of the coupling terms provides a direct identification to the original MHD equations from which the model is derived.

This scenario is in excellent agreement with our simulations, as illustrated in Fig.\ref{fig:lambdam} at $E=\infty$ (to be compared also with the results presented in Fig.\ref{fig:Rm_vs_Re}): the lower values for the dynamo threshold (characterised by $m_B=2$) are reached in the vicinity of the onset of the resonant mode (here $m_V=4$). The same mechanism is observed at smaller Ekman numbers, with again a 2:4 resonance at $E=10^{-2}$, and a 3:6 resonance at $E=10^{-3}$. The dynamos we investigated at $E=5.10^{-4}$ did not appear to lie in the vicinity of a wavy bifurcation, and we did not observe either the associated, sudden decrease in dynamo threshold that we find for other Ekman numbers. It is also remarkable that the dynamo selects a $m_B=1$ mode at $Re=150,E=\infty$ where the least stable velocity mode is the resonant mode $m_V=2=2m_B$. Our previous observation that the bistability region seems to tighten toward the TV onset is also well consistent with the recovery of a supercritical transition for $|\lambda | \gg 1$.

\begin{figure}[htb!]
\begin{center}
\includegraphics[width=0.5\textwidth]{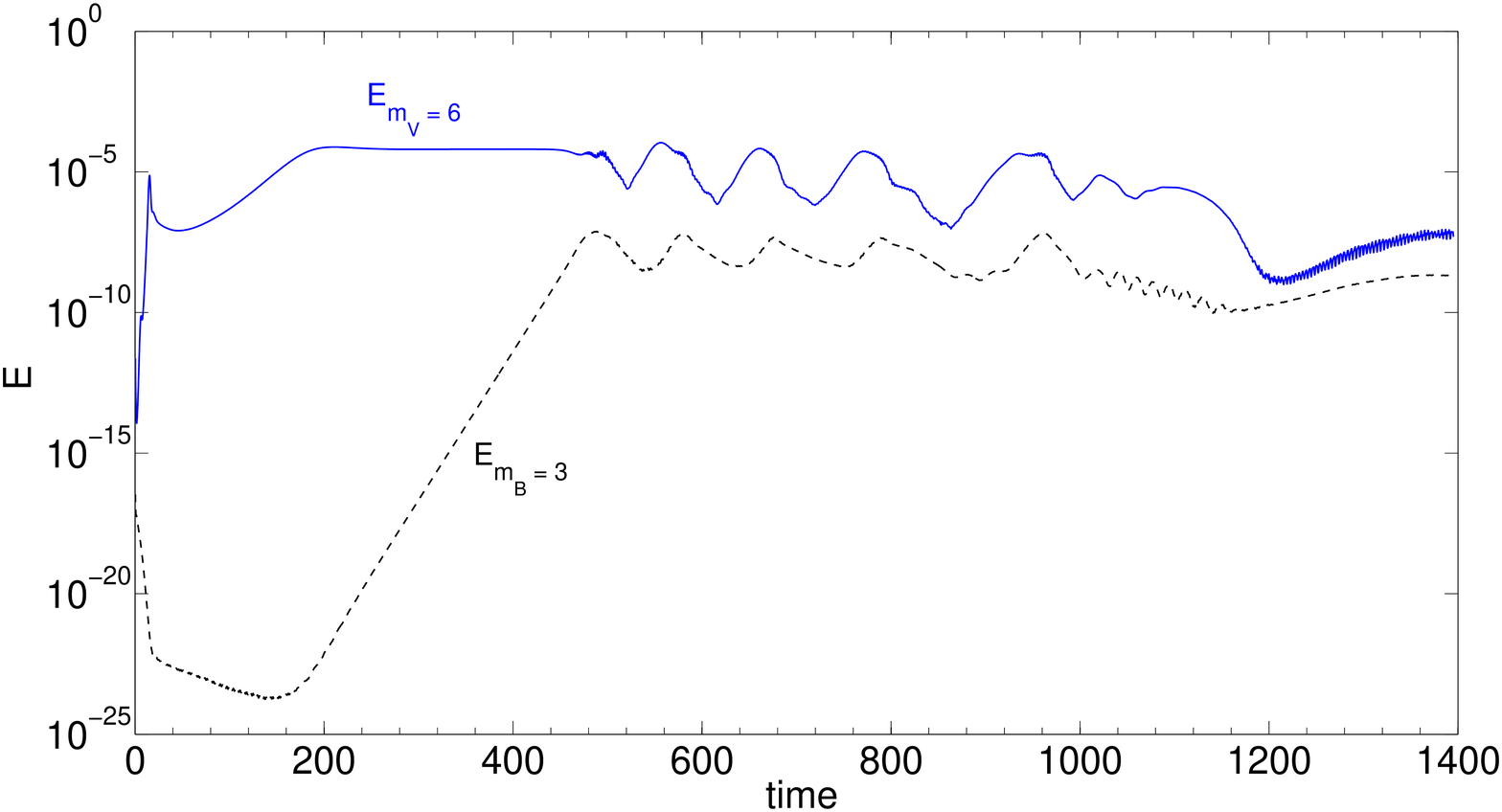}
\caption{Time-series of the  magnetic energy in the Fourier mode $m_B=3$  (dashed, black) and of the  velocity  mode $m_V=6$ (blue), for  $E=10^{-3}$, $Ta/Ta_c=11.92$, $Pm=1.25$. Both the dynamics and the structure of the dynamo are determined by the strong correlation between the two modes, as predicted by the model. }
\label{fig:Re3000temp}
\end{center}
\end{figure}

Another confirmation of this scenario is given by the time series shown in Fig.\ref{fig:Re3000temp} for $\{E=10^{-3};Ta/Ta_c=11.92;Pm=1.25\}$:  here, the system lies relatively far from the dynamo onset ($Pm=2.5Pm_c$)  and the magnetic field is characterised by several unstable Fourier modes. However in the kinematic phase, the dynamo clearly selects the magnetic mode ($m_B=3$) which is resonant with the hydrodynamically unstable mode $m_V=6$. The most unstable mode $m_B=3$ then rapidly saturates, and the non-linear phase is  characterised by coupled oscillations between these two resonant modes. Interestingly, this simulation eventually exhibits a secondary transition at large time, in which the magnetic field switches to  $m_B=2$, and is therefore associated to a bifurcation of the velocity field to the resonant mode $m_V=4$.

\section{Conclusions}

Previous studies have shown that laminar spherical Couette flow fails to sustain dynamo action, which can only build on a primary hydrodynamic instability of the equatorial jet or the Stewarston quasi-geostrophic layer. Although most of the interest in spherical dynamos is with convection-driven flows, we demonstrate here that centrifugal instability is a possible candidate for magnetic field amplification in a spherical shell. The toroidal, counter-rotating vortices of spherical Taylor-Couette flow maintain a non-axisymmetric dynamo in the equatorial region, at the Taylor-vortices scale. Without overall rotation of the system, the dynamo bifurcation is found to be subcritical and the critical magnetic Reynolds number seems to reach a constant value in the flow turbulent regime, where the magnetic field experiences intermittent bursts near the dynamo onset. Sufficiently far from the onset for centrifugal instability, the effect of overall rotation strongly reduces the dynamo threshold.
 
 This decrease of the dynamo onset together with the subcritical nature of the bifurcation is understood in the framework of a simple model: close to the hydrodynamic onset for wavy vortices, the Lorentz force produces a quadratic resonant interaction between the dynamo magnetic field of wavenumber $m$ and the velocity mode of wavenumber $2m$. In consequence, although the axisymmetric Taylor vortices control the magnetic field generation, the waviness of the flow can stabilize the dynamo solution well below its linear onset. Interestingly, this effect is observed even below the hydrodynamic onset of wavy vortices. When the wavy resonant mode is linearly unstable, the kinematic dynamo onset can also be significantly decreased.

It is now worth discussing the relevance of this scenario for real systems. Although it is known to play little role for the Earth's dynamo, differential rotation may be very important in other astrophysical objects. For instance, observations and theoretical models \cite{HGW15} suggest that surface zonal flows observed on the jovian planets can extend below the observable surface layers. Recent observations of stellar interiors have revealed, by means of asteroseismic techniques, the existence of  internal, radial differential rotation in red giants or subgiants (see e.g. \cite{D12},\cite{D14}), with a ratio of the core/envelope angular velocities in the 2.5-20 range.

If the angular momentum sufficiently decreases outwards near the equator in these different systems, the Taylor-vortex dynamo described in this paper may play a role in the generation of their magnetic field. This scenario may be particularly relevant for explaining magnetic fields that exhibit non-dipolar geometries, unlike those predicted for convection-driven flows.\\

 Such spherical Taylor-Couette dynamo also opens interesting perpectives for experimental flows: MHD spherical Couette experiments (\cite{Zimmerman14, Cabanes14}) are currently based on aspect-ratios similar to the Earth's core, thus forbidding the occurence of the centrifugal instability. Although the smallest $Rm$ reported in this paper ($Rm_{min}\sim150$) is still very challenging from an experimental point of view, one may imagine to use a smaller gap in these experiments, more prone to the generation of Taylor vortices in the equatorial plane. To that end, future work should focus on the search for $Rm$ smaller than the ones reported in this paper, by means of a more systematic study of this new spherical TC dynamo, particularly regarding the role of $Re$ and of the gap between the spheres.\\
Finally, the dynamical system (\ref{mode2m})-(\ref{modem}) describing the generation of a subcritical dynamo is mainly based on symmetry arguments and on the quadratic form of the Lorentz force. One can therefore expect it to be quite general and may be relevant beyond the present paper: it would be interesting to see if a similar dynamics occur in other systems, like for instance the MRI subcritical dynamo or convection-driven dynamo models of the Earth's core.

 \section{Ackowledgements}
This work was granted access to the HPC resources of MesoPSL financed by the Region Ile de France and the project Equip@Meso (reference ANR-10-EQPX-29-01) of the programme Investissements d'Avenir supervised by the Agence Nationale pour la Recherche.

\end{document}